\documentclass[12pt, preprint]{aastex}
\usepackage{amsmath,amsthm}

\newcommand\ts{\times}
\newcommand\beq{\begin{equation}}
\newcommand\eeq{\end{equation}}

\begin{document}

\shorttitle{Stochastic Wobble of Accretion Discs and Jets}
\shortauthors{Pettibone and Blackman}

\title{Stochastic Wobble of Accretion Discs and Jets from Turbulent  Rocket Torques}

\author{Ryan Pettibone$^{1,2}$ and Eric G. Blackman$^{1}$}

\affil{1. Department of Physics and Astronomy, University of Rochester, Rochester, NY 14627; 2. Present Address: Department of Physics, Caltech, Pasadena, CA, USA,
91125}


\begin{abstract}

Models of accretion discs and their associated outflows often incorporate assumptions of axisymmetry and  symmetry across the disc plane.  However, for turbulent discs these symmetries only apply to averaged quantities and do not 
apply locally.  The local asymmetries can induce  local imbalances in outflow power across the disc mid-plane, which can in turn 
induce local tilting torques. Here we calculate the effect of the resulting stochastic torques on disc annuli that are a consequence of standard mean field accretion disc models. The torques induce a random walk of the
vector  perpendicular to the plane of each averaged annulus. This random walk is characterized by a radially dependent diffusion coefficient which we calculate for small angle tilt. We  use the coefficient  
to calculate a radially dependent time scale for annular tilt and associated jet wobble. 
 The wobble time depends on the square of the wander angle so the age of a given system determines the maximum wobble angle. We apply this to 
examples of  blazars, young stellar objects and binary engines of pre-planetary nebulae and microquasars. It is noteworthy that for an averaging time  
$t_w\sim 3$ days, we estimate a wobble angle for jets in SS433 
of $\theta\sim 0.8$ degrees,
not inconsistent with  observational data. In general the  non-periodic nature of the stochastic wobble could distinguish it from faster periodic jet precession.

\end{abstract}


\section{Introduction}

Winds and jets are commonly associated with accretion disc engines, and 
standard models of discs \citep{ss73,pringle1981,kingfrankraine} 
and outflows \citep{bp82,pp92} typically  impose axisymmetry and reflection
symmetry across the disc plane.
However,  theory and observation suggest that for a wide range of sources, 
 angular momentum transport is likely mediated at least in part by turbulence 
(e.g. \cite{t02}). This immediately implies that the assumptions of axisymmetry
apply  at most only to mean flow quantities, averaged over sufficient 
time or space (e.g. Balbus et al. 1994).  The local violation of symmetries for a turbulent disc,  leads to an intrinsic variability (Blackman 1998).

We suggest that an additional 
consequence of the  fluctuations  may be  a local
``rocket effect.'' When a wind is expelled from the surface of a  disc 
it exerts a force on that surface which compensates for the momentum of the outflow.  In idealized models of symmetric accretion discs the forces exerted on the disc by the winds on either side of the disc cancel out.  
In reality, although these forces must cancel globally if the disc center of mass is stationary,  turbulence induced asymmetry makes it unlikely that these forces cancel out locally. Thus, different portions of the disc can be displaced above or below the initial  mean symmetry plane.
Here we investigate how disc turbulence can cause a local imbalance between the winds emanating from the top and bottom by averaging the local effect over azimuth. The net effect is to induce a stochastic wander of the vector perpendicular to the plane determined by the azimuthal average of the asymmetry
for each eddy at a given radius.  We model this as  Brownian motion, and calculate the associated diffusion coefficient.

In Sec. 2 we discuss  the needed basics of Brownian motion.  In Sec. 3 we then  derive the equation for the aforementioned stochastic  diffusion coefficient for tilt as a function of disc parameters. In Sec. 4 we apply this diffusion coefficient to the wobbling of a thin accretion disc  and a stochastic wander of the associated jet axis, assumed to be parallel to the disc's primary angular velocity vector. 
 We  apply the formalism to four astrophysical engine classes (blazars, young stellar objects (YSOs), planetary nebulae,  microquasars)
and estimate the wander angles for which the wobble could be observable
in these systems. As a particular example, 
we compare the crudely predicted wobble to that
observed in SS433 and find agreement to order of magnitude for a plausible
choice of disc parameters. The stochastic nature of the tilt distinguishes it from that of an ordered precession  
We conclude in Sec. 5.

\section{Review of Brownian Motion}

Consider the random walk 
on 
a two-dimensional square lattice of points \citep{ch43}.  We label a point on the lattice  by coordinates $(\theta,\phi)$, and the spacing between adjacent lattice points by $\Delta \theta$.  We also break time  into discrete intervals $\Delta t$.  We use  these coordinates in anticipation of our application to the random walk of a unit vector tip on the surface of a sphere, locally approximated as a plane for small angles. 

 At time $t-\Delta t$, a point particle located at $(\theta,\phi)$ moves by time $t$ to one of the 4 points $(\theta + \Delta \theta,\phi)$, $(\theta - \Delta \theta,\phi)$, $(\theta,\phi+\Delta \theta)$, $(\theta,\phi-\Delta \theta)$ with equal probability.  If the particle moves \emph{into} the lattice point $(\theta, \phi)$ at time $t$, it would  have come from one of the four adjacent lattice points in the previous time step.  Therefore, if $P(\theta, \phi, t)$ is the probability the particle is located at $(\theta, \phi)$ at time $t$, we have
\begin{equation}
\begin{array}{r}
P(\theta, \phi, t) = \frac{1}{4} (P(\theta+\Delta \theta,\phi,t-\Delta t) 
+P(\theta - \Delta \theta,\phi,t-\Delta t) \\ +  P(\theta,\phi+\Delta \theta,t-\Delta t) + P(\theta,\phi-\Delta \theta,t-\Delta t))
\end{array}
\label{1}
\end{equation}
This allows us to write a difference equation for $P(\theta, \phi, t)$.  Using (\ref{1}) we have
	\begin{equation}
	\frac{P(\theta, \phi, t)-P(\theta, \phi, t-\Delta t)}{\Delta t} = D
 \left(A		(\theta, \phi) + B(\theta, \phi) \right) 
\label{2}	
\end{equation}
where
	\begin{equation}
	A(\theta, \phi) = \frac{P(\theta + \Delta \theta,\phi,t-\Delta t) - 2P(\theta, \phi, t-\Delta t) + P(\theta - 		\Delta \theta,\phi,t-\Delta t)}{(\Delta \theta)^2},   
	\end{equation}

	\begin{equation}
	B(\theta, \phi) = \frac{P(\theta,\phi+\Delta \theta,t-\Delta t) - 2P(\theta, \phi, t-\Delta t)+ P(\theta,\phi-		\Delta \theta,t-\Delta t)}{(\Delta \theta)^2},
	\end{equation}
and
\begin{equation}
D=\frac{(\Delta \theta)^2}{4 \Delta t}.
\label{5}
\end{equation}
Equation (\ref{2}) is a discrete version of the two-dimensional diffusion equation that characterizes a random walk with diffusion coefficient $D$.  Although we will use the discrete version in this paper, a continuum diffusion
equation results by taking the continuum limit for  which $\Delta \theta \rightarrow 0$ and $\Delta t \rightarrow 0$.  In this limit, $A \rightarrow \frac{\partial^2 P}{\partial \theta^2}$ and $B \rightarrow \frac{\partial^2 P}{\partial \phi^2}$.  If we then define $D \equiv \lim_{\Delta t,\Delta \theta \to 0}(\Delta \theta)^2/ \Delta t$, we have:
\begin{equation}
	\frac{\partial P}{\partial t} =  D\left(\frac{\partial^2 P}{ \partial \theta^2}
+\frac{\partial^2 P}{ \partial\phi^2} \right)
\end{equation}
which is a two-dimensional Cartesian diffusion equation in the local angle coordinates.

\section{Derivation of the Wobble Diffusion Coefficient}

To derive $D$ for a disc, we first consider a small volume element at radius $r$ with azimuthal and radial scales  equal to that of a turbulent eddy and with vertical
scale equal to the disc thickness. 
 We define $dL$ as the net mechanical wind luminosity  emanating from this volume element, and $v$ as the associated mean outflow speed at the chosen radius.
We posit that the winds emanating from either side of thin annulus are not exactly in momentum balance at all times due to turbulent fluctuations.  We denote  wind mechanical luminosities for a given eddy emanating from the top and bottom of the disc as $dL_{top}$ and $dL_{bot}$ respectively, such that  $dL = dL_{top} + dL_{bot}$.  Let $\boldsymbol\epsilon_{top}$ and $\boldsymbol\epsilon_{bot}$ denote unit vectors in the direction of wind flow on the top and bottom.  The net force on the volume element from the two winds  is 
	\begin{equation}
	d\mathbf{F} = d\mathbf{F}_{top} + d\mathbf{F}_{bot} = \frac{dL_{top}}{v}\boldsymbol\epsilon_{top} + 	\frac{dL_{bot}}{v}\boldsymbol\epsilon_{bot} = \boldsymbol\epsilon' \frac{dL}{v} \mbox{,}
\label{6}
	\end{equation}
where $\boldsymbol\epsilon' \equiv (\boldsymbol\epsilon_{top}
 dL_{top}
+ \boldsymbol
\epsilon_{bot}dL_{bot}
)/dL$, and characterizes the direction of  wind imbalance.
Since we have assumed that the origin of the force imbalance (turbulence) is intrinsically random, $\boldsymbol\epsilon'$ will be a random vector dependent on both time and position. Fluctuations around both the mean outflow speed
and mass outflow rate can be incorporated into $\epsilon'$ given its definition; the explicitly presence of the mean velocity in (\ref{6}) does not preclude this.

The force on a disc volume element results from a ``rocket effect" created by  non-vanishing last term in (\ref{6}). 
To calculate the associated torque, we assume that $\boldsymbol\epsilon'$ is a poloidal vector ($\boldsymbol\epsilon' \cdot \hat{\phi} = 0$) for every volume element.
The associated torque  then lies in the initial disc plane and is given by 
	\begin{equation}
	d\boldsymbol\tau = (\mathbf{r} \times d\mathbf{F}) = (\mathbf{r} \times \boldsymbol\epsilon') \frac		{dL}{v} = (\hat{\mathbf{r}} \times \boldsymbol\epsilon') \frac{r}{v} dL \mbox{ .}
	\end{equation}
Summing  the torques for all volume elements at radius  $r$ leaves a net torque  on an annulus of radial thickness $dr$ at radius $r$ given by
	\begin{equation}
	d\boldsymbol\tau_{net} = \sum_{dV\in E(r)} (\hat{\mathbf{r}} \times \boldsymbol\epsilon') \frac{r}{v} 		dL
 = \boldsymbol\epsilon \frac{r}{v} \frac{dL}{dr} dr \mbox{ ,}
\label{9}	
\end{equation}
where we have defined $E(r) \equiv \{\mbox{all volume elements at radius r}\}$, and $\boldsymbol\epsilon\equiv \sum (\hat{\mathbf{r}} \times \boldsymbol\epsilon')
$. Note that since $\boldsymbol\epsilon'$ is poloidal, $\hat{\mathbf{r}} \times \boldsymbol\epsilon'$ lies in the plane of the disc, and therefore so does $\boldsymbol\epsilon$, being a sum of such vectors. 
Since $d{\boldsymbol\tau}$ lies in the plane of the disc, so does 
$d\boldsymbol\tau_{net}$.
By summing over all the volume elements in the annulus we have 
removed the azimuthal dependence, providing what is needed to 
 characterize the response of an 
axisymmetric mean field annulus. This is 
an annulus of a ``standard'' axisymmetric radially dependent accretion disc (which itself is axisymmetric precisely because of the  azimuthal average over turbulent fluctuations).

We can now calculate the rate of change of angular momentum of this annulus associated with  tilting from its initial disc symmetry plane. This
 tilt introduces an angular momentum  vector 
in the initial disc plane that bisects the annulus, and is  approximately perpendicular to the underlying Keplerian rotation vector.
The component  of the moment of inertia tensor appropriate for the tilt axis is $dI = {1\over 2}r^2 dm(r)$, where the  mass of the annulus is  $dm(r) = \mu(r)dr$ and $\mu(r)$ is the mass per unit radius.  The rate of change in angular momentum that characterizes the tilt is then 
\begin{equation}
d\dot{\mathbf{J}} = \frac{d\boldsymbol\omega}{dt} dI = \frac{r^2}{2} \frac{d\boldsymbol\omega}{dt} dm= \frac{r^2}{2} \frac{d\boldsymbol\omega}{dt} \mu(r) dr,  
\end{equation}
where $\boldsymbol\omega$ is the orbital angular momentum vector that would
reflect a fixed-plane orbital motion of disc material in the absence of rocket torques.
Using Newton's second law $d\dot{\mathbf{J}} = d\boldsymbol\tau_{net}$.  It follows that
\begin{equation}
\frac{r^2}{2} \frac{d\boldsymbol\omega}{dt} \mu(r) = \boldsymbol\epsilon \frac{r}{v} \frac{dL}{dr} \mbox{ .}
\label{10}
\end{equation}
If we assume that $\boldsymbol\epsilon$ varies on a characteristic time scale $\Delta t$, then (\ref{10}) implies
\begin{equation}
\Delta\boldsymbol\omega = \frac{2}{r v \mu(r)} \frac{dL}{dr} \boldsymbol\epsilon \Delta t.
\label{11}
\end{equation}
Eq. (\ref{11}) measures the tilt of the plane the 
 annulus.  Since $d\boldsymbol\tau_{net}$ is in the plane of the annulus, the torque has the effect of rotating $\boldsymbol\omega$ but not changing its magnitude.  If the angular deflection is small in a time $\Delta t $, then  
$ \Delta\theta \sim \frac{\left| \Delta\boldsymbol\omega \right|}{\omega} $, where $\Delta\theta$ is the change in the angle of $ \boldsymbol\omega$ during the time $\Delta t $.  We can then appeal to the Cartesian lattice 
caluclation of Sec. 2 to obtain the small angle diffusion coefficient:  
Combining the expression for $\Delta \theta$ just derived 
 with (\ref{11}) and  (\ref{5}) we obtain 
\begin{equation}
D(r) = 
\frac{1}{r^2 {\mu(r)}^2 v^2 \omega^2} {\left( \frac{dL}{dr} \right)}^2 \epsilon^2 \Delta t=
\frac{1}{r^2 {\mu(r)}^2 v^2 \omega^2} {\left( \frac{dL}{dr} \right)}^2 \frac{\epsilon^2_{ed}  t_{ed}}{ \sqrt N},
\label{13a}
\end{equation} 
where $\epsilon_{ed}=N^{1/2}\epsilon$
 represents the characteristic 
contribution from each of the $N$ eddies at a given radius
 to $\epsilon$, and $\tau_{ed}=\tau/N^{1/2}$ is the 
characteristic time scale for evolution of a given eddy.
Here  $N=2\pi r/l_{ed}$ is the number of  eddies with scale $l_{ed}$ at radius $r$.

This diffusion coefficient characterizes the random walk of the orbital velocity vector of the mean field annulus due to tilt from the stochastic rocket force.
Since the magnitude of $\boldsymbol\omega$ never changes in our approximation, the tilt implies that the orbital angular vector moves on the surface of a sphere as the annulus wobbles. For small tilt angles, this surface can be locally
approximated by a plane and we are justified 
in using the Cartesian formulation of Sec. 2, despite the global differences between spheres and planes.

The derivation and use of Eqn. (\ref{13a}) implies
that each radius  has its own wobble diffusion coefficient.  This does not violate angular momentum conservation because annuli are (at least) viscously coupled.  If all of the material
that enters the disc at large radii is coplanar, 
then conservation of total angular momentum 
implies that the total integrated angular momentum about any
axis perpendicular to that of the initial $\boldsymbol \omega$ must
vanish at all times. This means that if some inner mean annulus of gas tilts in one direction, there will be a corresponding counter-tilt elsewhere. 
Turbulent viscosity allows the radial distribution of tilts to  be non-trivial 
and because jet power is likely dominated by inner annuli, the tilt
of the inner-most  annuli are particularly relevant for predicting 
jet wobble because they dominate the accretion power. 
In what follows, we assume that the axis of any annular section of a jet
tracks the orbital axis of the corresponding disc annulus to which that jet section is anchored.  Thus the jet wobble 
directly tracks the disc wobble.
 

\section{Application to Astrophysical Engines}

Here we apply the above formulae  to thin  accretion discs and jet wobble.

\subsection{Determination of Stochastic Wobble times for Discs and Jets}

For a thin, non-self-gravitating disc in hydrostatic equilibrium
\begin{equation}
\omega = \omega_K \equiv \sqrt{GM/r^3} \sim  c_s/H \sim t_{ed}^{-1} 
\label{17}
\end{equation}
where $c_s$ is the sound speed,
$M$ is the central object mass,
 and $H$ is the density scale height.
The Shakura-Sunyaev (\cite{ss73})viscosity $\nu$ in  hydrostatic equilibrium then satisfies 
 $\nu \equiv \alpha c_s H \sim v_{ed}^2/ t_{ed} \sim v_{ed}^2H/c_s$,
where $\alpha$ is the viscosity parameter.
Then (e.g. Blackman 1998) $v_{ed}\sim \alpha^{1/2}c_s$ and 
\begin{equation}
l_{ed} = \alpha^{1/2} H,
\label{18}
\end{equation}
so 
\begin{equation}
N=\frac{2\pi}{ \alpha^{1/2}}\frac{r}{ H}.
\label{19a}
\end{equation}

The disc accretion rate can be modeled as
$\dot{M}_a(r) = \dot{M_{o}}(r/r_o)^s$
 (e.g. \cite{bb99}), where $\dot{M_{0}}$ is the accretion rate at the outer edge of the disc, $r_o$ is the radius of the outer edge, and $0< s< 1$ is a 
parameter.  
The total mass outflow rate from $r_o$ to $r$ is then
$\dot{M}(r) = \dot{M_o}[1-(r/r_o)^s]$. 
Taking the derivative, we obtain the
outflow mass loss rate by a small annulus of  width $dr$ at radius $r$ to be
\begin{equation}
\frac{d\dot{M}}{dr} dr = -\dot{M_o} \frac{s}{r_o} \left(\frac{r}{r_o}\right)^{s-1} dr . 
\label{14}
\end{equation}
The luminosity of the gas ejected by the wind in the annulus is $dL =  \frac{1}{2} {v}^2 |d\dot{M}|$, where the outflow speed satisfies 
\begin{equation}
v\simeq v_{esc}\equiv \sqrt{2GM/r},
\label{14a}
\end{equation}
so in combination with  (\ref{14}) we have
\begin{equation}
\frac{dL}{dr} = \dot{M_o} \frac{s}{2 r_o} \left(\frac{r}{r_o}\right)^{s-1} {v_{esc}}^2. 
\label{15}
\end{equation}



Using (\ref{17}), (\ref{18}), (\ref{19a}), (\ref{14a}) and (\ref{15}) 
 in 
(\ref{13a}) gives
\begin{equation}
\begin{array}{r}
D(r)
= 
 4.4\times 10^{19}\alpha^{1/4}\left({s\epsilon_{ed}\over \mu}\right)^2\left( {H\over r_g}\right)^{1\over 2}
\left({M\over M_\sun}\right)^{-1}\left({r\over r_0}\right)^{2s-1}
\left({r_0\over  r_g}\right)^{-1}
 \left({{\dot M}_0\over 10^{18} \ {\rm g/s}}\right)^2\ {\rm rad^2/ s^{}},
\label{19}
\end{array}
\end{equation}
where $r_g=GM/c^2$.
For a standard Shakura-Sunyaev disc model \citep{ss73} supplemented by our radially dependent
accretion rate ${\dot M}={\dot M}_o(r/r_o)^s$  we have
\begin{equation}
\mu(r) =
 5.1\times 10^{11} \alpha^{-4/5} \left({\dot M}_0\over 10^{18}{\rm g}/{\rm s}\right)^{7/10} 
\left(\frac{r}{r_o}\right)^{7s/10} 
\left({M\over M_\odot}\right)^{1/2} 
\left({r \over r_g }\right)^{1/4}\left[1-\left({r_g\over r }\right)^{1/2}\right
]^{7/20}
\ {\rm g/cm},
\label{20}
\end{equation}
and
\begin{equation}
{H\over r_g} = 9\times 10^{-3}\alpha^{-1/10}\left({M\over M_\odot}\right)^{-1/4}
 \left({{\dot M}_0\over 10^{18} \ {\rm g/s}}\right)^{3/20}
\left({r\over r_o}\right)^{3s/20}\left({r\over r_g}\right)^{9/8}
\left[1-\left({r_g\over r }\right)^{1/2}\right]^{3/20}.
\label{21}
\end{equation}
Using (\ref{20}) and (\ref{21})
 in (\ref{19}), 
the time required to wander an angle $\theta$ radians is 
\begin{equation}
\begin{array}{r}
t_{w}(r)= 
 {\theta^2\over 4D} = 1.6\times 10^4 
 \alpha^{-9/5} \left({\theta\over s\epsilon_{ed}}\right)^{2}\left({\dot M}_0\over 10^{18}{\rm g}/{\rm s}\right)^{-27\over 40} 
\left(\frac{r}{r_g}\right)^{{17\over 16}-{27s\over 40}} 
\left({M\over M_\odot}\right)^{17/8} 
\left({r_0 \over r_g }\right)^{27s/40}\left[1-\left({r_g\over r }\right)^{1/2}\right
]^{5/8}
\ {\rm sec}.
\label{23}
\end{array}
\end{equation}

For stochastic wobble to be observable $t_{w}(\theta)$ 
must at least be less than the age of a system $\tau_{age}$. 
A further constraint follows from  the standard accretion  paradigm 
in which accretion supplies the energy for a turbulent viscosity 
which, together with the outflow, transport angular momentum and sustain the accretion. 
As discussed at the end of Sec.3, the turbulent viscosity is also a means of 
transporting the angular momentum associated with 
tilt between annuli such that
the tilt of inner annuli is compensated by the opposing tilt of 
outer annuli.  For stochastic wobble to produce a net mean tilt
on an annulus at an inner radius, 
the viscous time scale there $\tau_\nu(r)$  should be short
compared to the tilt time, so that the compensating 
angular momentum opposite to that associated with the tilt 
can be transported outward.
Note that even though material also accretes on a viscous time,
the resupply of material at a given radius would also maintain the stochastic 
equilibrium on time scales short compared to the secular wobble time.
Provided there is a supply of mass, we therefore consider  
viscosity to maintain the conditions for 
the effect to accrue, rather than washing out the
effect at a given radius.

Summarizing the above conditions to be checked, we have
\beq
\tau_v (r) < t_{w}(r) < \tau_{age}, 
\label{23b}
\eeq
where, from (\ref{21}),   the viscous accretion time scale for $r > 3 r_g$ is given by
\begin{equation}
\tau_v(r)\simeq 
{r^2\over \alpha H^2\Omega}
= 6\times 10^{-2} 
\alpha^{-4/5}
\left({M\over M_\odot}\right)^{3/2}
 \left({{\dot M}_0\over 10^{18} \ {\rm g/s}}\right)^{-3/10}
\left({r\over r_g}\right)^{-{3s\over 10}+{5\over 4}}\left({r_0\over r_g}\right)^{3s/10}
{\rm sec},
\label{24}
\end{equation}
and 
\begin{equation}
\tau_{age}=  Max\left[\tau_v(r_o), {M_{c}\over {\dot M}}\right]
\label{ineq}
\end{equation}
where $M_c$ is the companion mass. The second value in brackets
is used when the accretion disc is being supplied by a companion, while the
first (the viscous time at the outer disc radius) is used when there is no
companion-fed resupply.

For later use in evaluating (\ref{23b}), 
we have from (\ref{23}) and (\ref{24}) 
\begin{equation}
\begin{array}{r}
{t_{w}(r)\over \tau_\nu(r)}\sim 
 2.7\times 10^5 
 \alpha^{-1} \left({\theta\over s\epsilon_{ed}}\right)^2\left({\dot M}_0\over 10^{18}{\rm g}/{\rm s}\right)^{-3/8} 
\left(\frac{r}{r_g}\right)^{-{3\over 16}-{3s\over 8}} 
\left({M\over M_\odot}\right)^{5/8} 
\left({r_0 \over r_g }\right)^{3s/8}
\label{24c}
\end{array}
\end{equation}
and
\begin{equation}
\begin{array}{r}
{t_{w}(r)\over \tau_\nu(r_o)}\sim 
 2.7\times 10^5 
 \alpha^{-1} \left({\theta\over s\epsilon_{ed}}\right)^2\left({\dot M}_0\over 10^{18}{\rm g}/{\rm s}\right)^{-3/8} 
\left(\frac{r}{r_g}\right)^{{17\over 16}-{27s\over 40}} 
\left({M\over M_\odot}\right)^{5/8} 
\left({r_0 \over r_g }\right)^{{27s\over 40}-{5 \over 4}}
\label{24cc}
\end{array}
\end{equation}
Note that the left side of (\ref{24cc}) employs $r$ for the numerator and $r_o$ in the denominator on the left hand side as that ratio represents a  
comparison between  the wobble time at a given radius
with the accretion time at the outer radius.
Eq. (\ref{24c}) compares the two time scales at the same radius.
In the next section we estimate (\ref{23}) and check
condition (\ref{23b}) for example classes of  accretion engines.

\subsection{Application to Astrophysical Jets}


To find the characteristic jet/disc wobble times for various systems, 
we now estimate (\ref{23})  for  active galactic nuclei (AGN) 
 young stellar objects (YSO), 
microquasars, planetary nebulae (PNe) and active galactic nuclei (AGN).
Since the wander time scale depends on the square of the wander angle, 
we can determine the wander angle for which 
the observability condition (\ref{24}) is satisfied. 
Presently, the quantities $s$ and $\epsilon_{ed}$, and $\alpha$
are not tightly  
pinned by either theory or simulation so we maintain their dependence
explicitly. Fiducial values will be taken to be 
$\epsilon_{ed}=0.5$, and a mass-loss power law index $s=0.2$, so their
product can be scaled to $0.01$.
We also  scale the results with $\alpha$.

Consider first AGN blazars (e.g. 3C273;  Paltani and T\"urler 2005).
We use  $M\sim 10^9M_\odot$ and ${\dot M}_0\sim 10 M_\odot
/{\rm yr}$. Then taking  $r_0=10^6 r_g$ and $r=r_g\sim 10^{14}$cm, 
we find that 
$t_{w}\sim 
3.1\times 10^{11}\left({s\epsilon_{ed}\over 0.01}\right)^{-2}\left({\alpha\over 0.1}\right)^{-9/5}
\left({\theta\over 0.03 \ {\rm rad}}\right)^2$ yr.
For these parameters the condition $\theta <0.03$ corresponds
to satisfying the second inequality in (\ref{23b}), namely that 
Eq. (\ref{24cc}) is $\le 1$. However, observable duty cycles of 
AGN jets seem to last only $\sim 10^8$ yr, which would requires 
a wobble angle  $\theta<100$''.
For this latter range of wobble angles, both inequalities in (\ref{23b}) are satisfied, using the first value on the right of (\ref{ineq}).


For  YSOs (Hartmann 1998), we take
$M= 1M_\odot$, $\dot M\sim 10^{-5}M_\sun$/yr, $r_o=100$ AU.
At  $r=7\times 10^7r_g=10^{13}$cm, 
Eq. (\ref{23}) then gives 
$t_{w}
\sim 4.6\left({s\epsilon_{ed}\over 0.01}\right)^{-2}\left({\alpha\over 0.3}\right)^{-9/5}\left({\theta\over 0.03 
 {\rm rad}}\right)^2$ yr. 
Using (\ref{24c}), the first inequality of (\ref{23b}) is easily satisfied.
For the second inequality of (\ref{23b}) we use the first value on the right of
(\ref{ineq}). From (\ref{24cc}) we then find  
${t_w(r)\over \tau_v(r_o)}
\sim 0.3 \left({\alpha\over .3}\right)^{-1}\left({\theta\over 0.01}\right)^{2}$ 
satisfied for wander angles $\le 0.01$rad.

Accretors with binary companions supply
accretion for times much longer than the viscous time scale at the outer
radius. This makes the second equality in (\ref{23b}) easier to satisfy for large wobble angles in those systems. Two examples of such are accretors are 
pre-PNe engines and microquasars.
For pre-PNe engines, an accretion disc can form around the
post-ascending giant branch (AGB) white dwarf core as the result of 
 common envelope evolution (Nordhaus \& Blackman 2006). 
We take 
$M= 0.6M_\odot$, $\dot M\sim 10^{-5}M_\sun$/yr, $r_o=10^{11}$cm.
Here the  engine accretor is a proto-white dwarf of radius 
$\sim  10^9$ cm or $r \sim 10^4 r_g$. 
At $r=10^5 r_g\sim 10^{10}$cm, 
Eq. (\ref{23}) then gives 
$t_{w}
\sim 5.5 \left({s\epsilon_{ed}\over 0.01}\right)^{-2}\left({\alpha\over 0.3}\right)^{-9/5}\left({\theta\over 0.03 \ {\rm rad}}\right)^2$ 
yr. 
Again for this case the first inequality in (\ref{23b}) is satisfied from
(\ref{24c}): At the inner disc radius
 $t_w/\tau_{\nu}\sim 700 \left({\theta\over 0.03 \ {\rm rad}}\right)^2$.
In this case however, the second equality in (\ref{23b}) 
is more easily satisfied as we can use the second value in (\ref{ineq}): 
The  pre-PNe outflows last $\tau_{age}\sim 10^3$ yr 
and the same accretion disc, supplied by a very low mass star, or
very high mass planet, may in fact
be continuously supplying outflow power even during the
$10^4{\rm yr}$ PNe phase (Blackman et al. 2001).
Both inequalities of (\ref{23b}) can then be satisfied for $\theta \le 0.8$ rad.

For  microquasar examples we consider  Cygnus X-1 (Young et al. 2001) and
SS433 (Katz \& Piran 1982; Begelman et al. 2006; 
Blundell et al. 2005;2007). For the former, we  take 
$M= 10M_\odot$, $\dot M\sim 10^{19} M_\sun$/yr, and $r_o=1000 r_g$.
At $r=r_g\sim 1.5 \times 10^6$cm, 
Eq. (\ref{23}) then gives
$t_{w}\sim 0.03\left({s\epsilon_{ed}\over 0.01}\right)^{-2}\left({\alpha\over 0.3}\right)^{-9/5}\left({\theta\over 0.03 \ {\rm rad}}\right)^2$ yr. 
For this case too, the first inequality in (\ref{23b}) is easily 
satisfied from
(\ref{24c}).  The second inequality in (\ref{23b}) allows
use of the second value in (\ref{ineq}). The companion mass to e.g. Cyg X-1
is at least 20$M_\sun$ so for accretion at $10^{-7}M_\odot$/yr, accretion can be powered long after the wobble time. 

For SS433 (e.g. Begelman et al. 2006)
$M=10M_\odot$, and we take the inner radius $r\sim r_g\sim 1.5 \ts 10^{6}$cm.
The accretion rate that makes it to the inner region, relevant for the jets, is 
${\dot M}_{in}\sim  2\times 10^{19}$g/s. This is to be distinguished from
the super-Eddington accretion rate of ${\dot M}_{out}\ge 10^3 {\dot M}_{in}$ at the disc circularization radius of 
$r_{circ}\sim 10^{12}$cm.  Begelman et al. (2006) argue that the much of the mass is taken out by a wind at $r\sim{r_{circ}\over 10}\sim 10^5r_g\sim 10^{11}$cm. 
We therefore take $r_o\sim 10^5r_g$ for the outer radius of the disc supplying
mass to the inner jet and use ${\dot M}_{in}$ for the relevant accretion rate.
We then obtain 
$t_{w}\sim 0.04\left({s\epsilon_{ed}\over 0.01}\right)^{-2}\left({\alpha\over 0.3}\right)^{-9/5}\left({\theta\over 0.03 \ {\rm rad}}\right)^2$ yr. 
For SS433, like Cygnus X-1, the first inequality in (\ref{23b}) is easily 
satisfied from (\ref{24c}) and a 
companion mass of $\sim 20M_\odot$
(based on crudely averaging recent mass estimates of Lopez et al. 2006
and Hillwig and Gies 2008)
allows use of the second value in (\ref{ineq}) for the second inequality in (\ref{23b}). This implies that accretion will be fueled long after the 
wobble time. Again, we have not considered relativistic effects.

We can compare the stochastic wander just estimated for SS433
 with the data of   Blundell et al. (2005,2007), 
who monitored fluctuations in the speed and direction of the inner jets.  
Table 1 of Blundell et al. (2007) shows that when averaged over
several day time scales, wander angles of $\sim 0.5-0.7$ degrees were obtained.
This is 
roughly consistent with our result above: Converting units,
out result for SS433 can be written
$\theta\sim 1.7\left({t_w\over  15 {\rm days}}\right)^{1/2}$ deg.

 There is likely a maximum time scale 
$t_{w,max}$ which is the  value of $t_w$
above which the stochastic wander is tempered by some negative feedback from
forces not considered in the present work.
Thus  $t_w\le t_{w,max}$ if the wander is to be calculated with our approximations.
One limit on $t_w$ 
 comes from the small wander angle approximation that allowed our use of 
 a 2-D Cartesian approximation. 
 However, for a precessing and nodding system like SS433  (e.g. Fabrika 2004), 
the 162 day precession time scale or the  6.3 day nodding time scale
 may provide the  maxium $t_w$ appropriate for computing  untempered stochastic wander. 
We do not determine $t_{w,max}$ here or study large angle wobble
 but note that a persistent feature of  stochastic wander 
 is its  non-periodicity, which distinguishes
it from  systematic precession induced by binary companions 
(e.g. Terquem et al. 1999).


\section{Conclusions}

Turbulence in accretion discs violates local axisymmetry and reflection
symmetry even if these properties hold in some average mean sense.  
As such, this symmetry violation likely also applies to outflows 
emanating from the disc.  
Due to a rocket-like effect, any outflow power asymmetry locally displaces the 
disc material in the direction opposite to the net outflow.  By azimuthally averaging  outflows with locally 
reflection asymmetric powers, we have calculated the radially dependent 
small-angle diffusion coefficient of the angular momentum vector for  
disc annuli.  The diffusion of this vector represents a stochastic 
tilting of the annulus defined by the averaging, and implies a disc warping
and jet wobble. For the latter, the most relevant tilt is that associated  inner disc radii. We  have calculated estimated the 
stochastic wobble angle vs. time scale for jets 
and applied the relation to a handful of astrophysical engine classes 
to exemplify crude application of the paradigm. The numbers
are not inconsistent with measured values in SS433.
We have not considered  the nonlinear saturation of the stochastic wobble
in the present work.



\acknowledgments EGB acknowledges support from  NSF grant AST-0406799, NASA grant ATP04-0000-0016, and NORDITA  during  the Turbulence and Dynamos workshop 2008.


\begin{thebibliography}

\bibitem[Balbus et al.(1994)]{bgh94} Balbus, S.~A., Gammie, 
C.~F., \& Hawley, J.~F.\ 1994, \mnras, 271, 197 


\bibitem[Begelman et al.(2006)]{2006MNRAS.370..399B} Begelman, M.~C., King, 
A.~R., \& Pringle, J.~E.\ 2006, \mnras, 370, 399 


\bibitem[Blandford \& Begelman(1999)]{bb99} Blandford, R. D., Begelman, M. C., 1999, \mnras, 303, L1
\bibitem[Blandford \& Payne(1982)]{bp82} Blandford, R. D., Payne, D. G., 1982, \mnras, 199, 883
\bibitem[Blackman et al.(2001)]{2001ApJ...546..288B} Blackman, E.~G., Frank, A., \& Welch, C.\ 2001, \apj, 546, 288 
\bibitem[Blackman(1998)]{1998MNRAS.299L..48B} Blackman, E.~G.\ 1998, \mnras, 299, L48 

\bibitem[Blundell 
\& Bowler(2005)]{2005ApJ...622L.129B} Blundell, K.~M., \& Bowler, M.~G.\ 2005, \apjl, 622, L129 


\bibitem[Blundell et 
al.(2007)]{2007A&A...474..903B} Blundell, K.~M., Bowler, M.~G., \& Schmidtobreick, L.\ 2007, \aap, 474, 903 

\bibitem[Chandrasekhar(1943)]{ch43} Chandrasekhar, S., 1943, Rev. Mod. Phys. 15, 1

\bibitem[Collins 
\& Newsom(1986)]{1986ApJ...308..144C} Collins, G.~W., II, \& Newsom, G.~H.\ 1986, \apj, 308, 144 


\bibitem[Collins 
\& Scher(2002)]{2002MNRAS.336.1011C} Collins, G.~W., \& Scher, R.~W.\ 2002, \mnras, 336, 1011 

\bibitem[Eikenberry et al.(2001)]{2001ApJ...561.1027E} Eikenberry, S.~S., 
Cameron, P.~B., Fierce, B.~W., Kull, D.~M., Dror, D.~H., Houck, J.~R., 
\& Margon, B.\ 2001, \apj, 561, 1027 

\bibitem[Fabrika(2004)]{2004ASPRv..12....1F} Fabrika, S.\ 2004, 
Astrophysics and Space Physics Reviews, 12, 1 


\bibitem[Fender et al.(2006)]{2006MNRAS.369..603F} Fender, R.~P., Stirling, A.~M., Spencer, R.~E., Brown, I., Pooley, G.~G., Muxlow, T.~W.~B., \& Miller-Jones, J.~C.~A.\ 2006, \mnras, 369, 603 
\bibitem[Frank et al.(2002)]{kingfrankraine} Frank, J., King, A., 
\& Raine, D.~J.\ 2002, Accretion Power in Astrophysics, by Juhan Frank and Andre
w King and Derek Raine, pp.~398.~ISBN 0521620538.~Cambridge, UK: Cambridge Unive
rsity Press, February 2002.,
\bibitem[Hartmann(1998)]{h98}Hartmann, L., {\it Accretion Processes in Star Formation}, 1998,  (Cambridge Univ. Press: Cambridge UK)

\bibitem[Hillwig 
\& Gies(2008)]{2008ApJ...676L..37H} Hillwig, T.~C., \& Gies, D.~R.\ 2008, \apjl, 676, L37 

\bibitem[Lopez et al.(2006)]{2006ApJ...650..338L} Lopez, L.~A., Marshall, 
H.~L., Canizares, C.~R., Schulz, N.~S., 
\& Kane, J.~F.\ 2006, \apj, 650, 338 

\bibitem[Katz 
\& Piran(1982)]{1982ApL....23...11K} Katz, J.~I., \& Piran, T.\ 1982, \aplett, 23, 11 


\bibitem[Margon 
\& Anderson(1989)]{1989ApJ...347..448M} Margon, B., \& Anderson, S.~F.\ 1989, \apj, 347, 448 

\bibitem[Nordhaus \& Blackman(2006)]{2006MNRAS.370.2004N} Nordhaus, J., \& Blackman, E.~G.\ 2006, \mnras, 370, 2004 
\bibitem[Paltani \& T\"urler(2005)]{2005A&A...435..811P} Paltani, S., T\"urler, M.\ 2005, \aap, 435, 811 
\bibitem[Pelletier \& Pudritz(1992)]{pp92} Pelletier, G., \& Pudritz, R.~E.\ 1992, \apj, 394, 117 
\bibitem[Pringle(1981)]{pringle1981} Pringle, J.~E.\ 1981, \araa, 19, 137 
\bibitem[Shakura \& Sunyaev(1973)]{ss73} Shakura, N. I., Sunyaev, R. A., 1973, A\&A, 24, 337
\bibitem[Terquem et al.(1999)]{t99}Terquem, C., Eisl{\"o}ffel, J., Papaloizou, J.~C.~B., 
\& Nelson, R.~P.\ 1999, ApJL, 512, L131
\bibitem[Terquem (2002)]{t02} Terquem, C.E.J.M.L.J,
2002, EAS Publications Series, 3, 203 

\bibitem[Young et al.(2001)]{2001MNRAS.325.1045Y} Young, A.~J., Fabian, 
A.~C., Ross, R.~R., \& Tanaka, Y.\ 2001, \mnras, 325, 1045 
\bibitem[Yuan, Markoff,\& Falcke(2002)]{ymf02} Yuan, F., Markoff, S., Falcke, H., 2002, A\&A, 383, 854


\end{thebibliography}
\end{document}